\RequirePackage{fix-cm}
\documentclass[twocolumn]{svjour3}

\usepackage{epsfig}

\journalname{Granular Matter}

\def\amc{\alpha_\mathrm{MC}}
\def\eyy{\dot\varepsilon_{yy}}
\def\exx{\dot\varepsilon_{xx}}
\def\syy{\sigma_{yy}}
\def\sxx{\sigma_{xx}}

\begin{document}

\title{Minimal dissipation theory and shear bands in biaxial tests}

\author{Thomas Stegmann \and J\'anos T\"or\"ok \and Lothar Brendel
\and Dietrich E. Wolf}

\institute{Faculty of Physics and CeNIDE, University of
Duisburg-Essen, 47048 Duisburg, Germany}

\date{Received: date / Accepted: date}

\maketitle

\begin{abstract}
True biaxial tests of granular materials are investigated by applying
the principle of minimal dissipation and comparing to two dimensional
contact dynamics simulations. It is shown that the macroscopic steady
state manifested by constant stress ratio and constant volume is the
result of the ever changing microscopic structure which minimizes the
dissipation rate. The shear band angle in the varying shear band
structures is found to be constant. We also show that introducing
friction on the walls reduces the degeneracy of the optimal shear band
structures to one for a wide range of parameters which gives a
non-constant stress ratio curve with varying aspect ratio that can be
calculated.
\end{abstract}

\keywords{biaxial shear, granular flow, shear band, least
 dissipation, optimization, Mohr-Coulomb theory}

\section{Introduction}

Shear bands seem inevitable whenever granular material is subject to
quasistatic deformation. The form and the position of these failure
zones depend on the experimental setup \cite{Nedderman} and sometimes
even on the specific sample \cite{Desrues84}.

Classical continuum theories predict strain fields from the stresses
of the material under load. Generally accepted constitutive relations
are still missing for dense granular materials, not to mention a
microscopic understanding of the parameters.

Therefore it is legitimate to approach the physics of shear bands from
a radically simplified perspective in which one idealizes shear bands
as strain field singularities and regards them as the basic dynamical
quantities of the theory. This is the approach used in this paper
where shear bands are dividing the granular material into domains
which are rigidly displaced with respect to each other. The realized
shear band configuration is determined using the principle of least
dissipation \cite{Unger04,Unger07}.

We apply this method to the {\em true biaxial tester}
\cite{Kadau06,Schwedes} containing a non-cohesive granular material,
where steady state flow can be observed. It allows the deformation of
a granular medium in a brick shaped container with perpendicular,
frictionless walls. This keeps the principal axes of the stress and
strain tensors constant and aligned perpendicular to the container
walls. We would like to find out what is the structure of the flow
that allows for an apparent macroscopic steady state, and under what
conditions can one predict the velocity field of the sample.

In this paper we present first some numerical simulation results on
true biaxial tests to motivate the theoretical model, which is then
solved for the setup with frictionless and later with frictional
walls. Finally we test the predictions on the simulation
results.

\section{Simulation results}

\subsection{Definition of the system}

In this paper we use distinct element simulations to study the
shearing of frictional granular material in the so called {\em true
biaxial tester} \cite{Kadau06,Schwedes}, which consists of two fixed
walls normal to the $z$ direction, stress controlled confining walls
normal to the $x$ direction (normal stress $\sigma_{xx}$), and strain
controlled compressing walls perpendicular to the $y$ direction
(normal strain rate $\dot\epsilon_{yy}{<}0$).

\begin{figure*}
\epsfig{figure=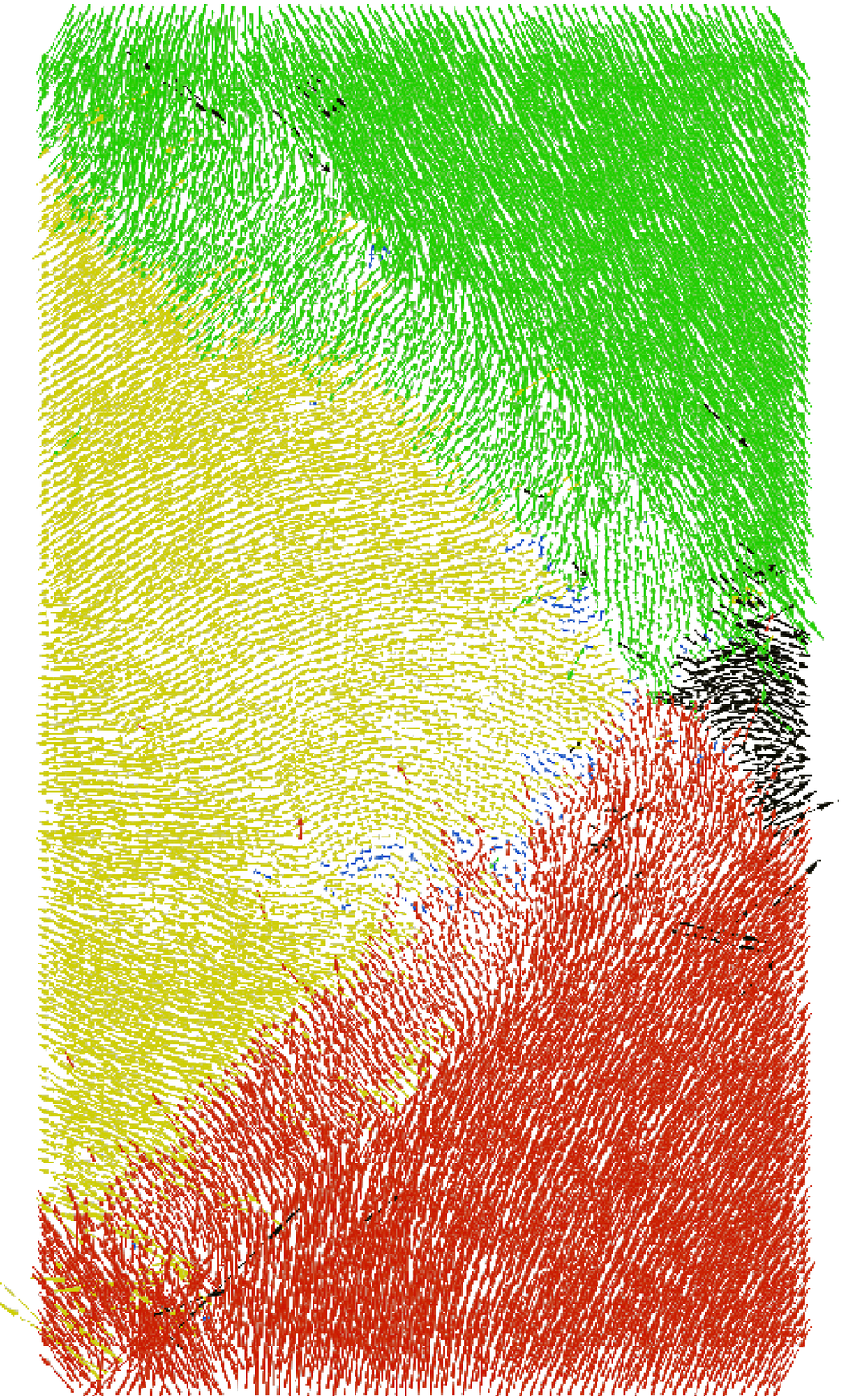,height=7.8cm}\raisebox{2ex}{(a)}
\epsfig{figure=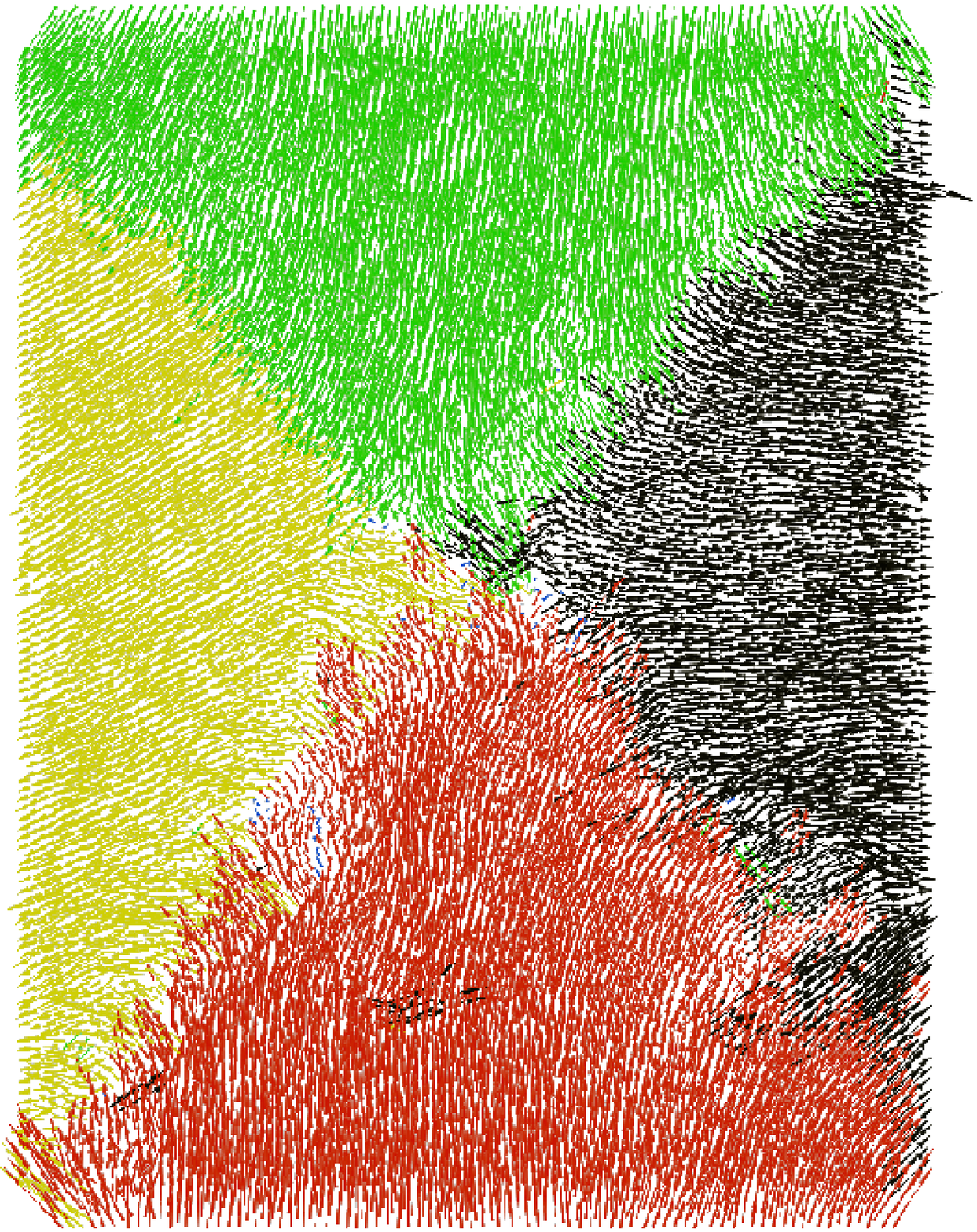,height=6.7cm}\raisebox{2ex}{(b)}
\epsfig{figure=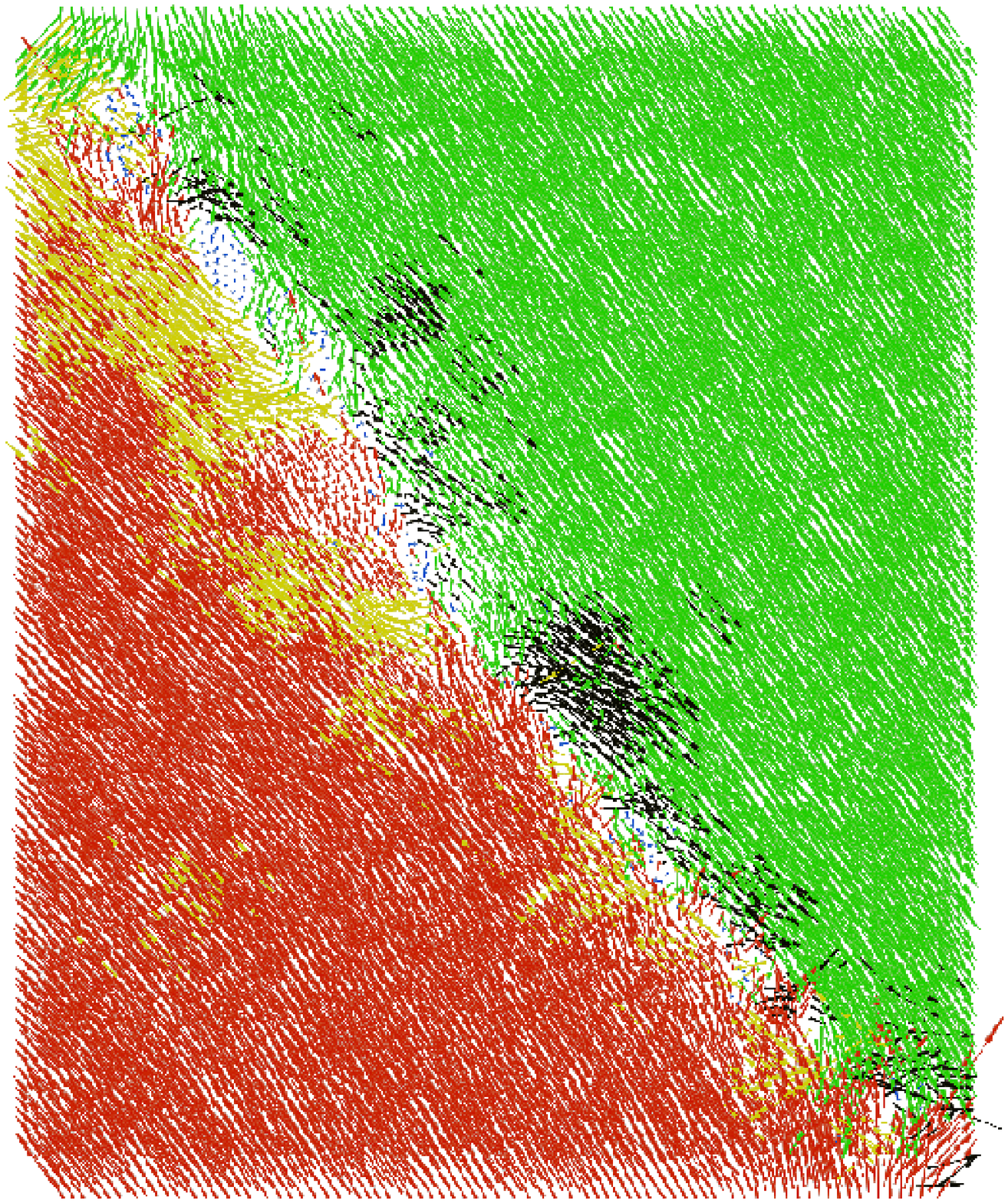,height=6.5cm}\raisebox{2ex}{(c)}
\caption{\label{Fig_SBex} Snapshots from simulated biaxial tests. Time
is progressing from (a) to (c). The particles are represented with
arrows proportional to their actual velocity. The color (online) is given
according to which quadrant the vector is pointing. Courtesy of D.
Kadau \cite{Kadau06}.}
\end{figure*}

The simulations were done using the contact dynamics method \cite{CD}.
Since the behavior is invariant along the $z$ direction we simulate
only a two-dimensional system in the $xy$ plane. Samples of 10000
frictional discs were enclosed in a rectangle by four walls. For the
constant confining stress $\sxx$ on the vertical and compressing
strain rate $\eyy$ on the horizontal walls we chose values such that
the inertial number \cite{daCruz} was always of the order of $10^{-4}$
\cite{Kadau06} so that the deformation can be regarded as quasistatic.
Further details of the simulations can be found in \cite{Kadau06}. The
particles were frictional with microscopic friction coefficient
$\mu_\mathrm{mic}{=}0.3$. Tests with different particle numbers and friction
coefficients show qualitatively identical results, and thus are not
presented in this paper.

First, we present results with {\em frictionless walls}, corresponding
to the usual experimental setup \cite{Schwedes}.

\subsection{Shear band patterns}
\label{Sec_Shearbandpatterns}

For frictionless walls, after an initial transient, the stress $\syy$
measured on the strain controlled and the strain rate $\exx$ on the
stress controlled walls fluctuate around a constant value
\cite{Kadau06}. This was interpreted as a steady state, however, the
underlying microscopic behavior remains unknown.

On Fig.~\ref{Fig_SBex} we show some instantaneous velocity snapshots
of a simulation at different instances. Visual observation of the
velocity snapshots clearly indicate the existence of shear bands which
are narrow straight lines reflected by the boundary or absorbed by the
corners. In a different, two dimensional setup Hall {\it et al.} also
reported complicated shear band structures~\cite{Hall10}.

Simulations with different initial configurations show in general
different shear band structures at given instances, however certain
easily recognizable configurations appear around a given aspect ratio
and never too far from it.

The above observations give rise to two important questions which we
want to answer in this paper: (i) Why does the stress ratio
$\syy/\sxx$ remain constant in spite of the strongly varying velocity
field, (ii) Is it possible to predict the shear band patterns in
biaxial tests? In the next section we try to answer these questions
based on the principle of minimal dissipation.

\section{Principle of minimal dissipation}

\subsection{Theory}

Before introducing the principle of minimal dissipation we summarize
our assumptions: The system is described in the frame, where the
center of mass is fixed at the origin. We assume that the sample has
already reached {\it steady state} flow \cite{Pena06} and is
quasistatic, i.e.  there are no accelerations. Ignoring the small
volume fluctuations allows the calculation of the velocities of all
walls from the given strain rate $\dot\epsilon_{yy}$: The upper and
lower walls move with velocity $-u_y$, respectively $u_y$, the left
and right ones with $-u_x$, respectively $u_x$, where

\begin{equation}
u_y=L_y|\dot\epsilon_{yy}|/2 \quad \mathrm{and} \quad 
u_x=L_x\dot\epsilon_{xx}/2
   =L_x|\dot\epsilon_{yy}|/2.
\end{equation}

We model the shear bands as infinitely narrow, piecewise straight
lines running all the way from one container wall to another without
splitting or merging. These lines cut the system into domains inside
which all grain velocities are the same. The domain boundaries are
either shear band segments or walls.

The velocity of the particles in a domain must fulfill the following
boundary conditions: (i)~At a wall the normal component of a domain
velocity must coincide with the velocity of the wall. The parallel
component of the domain velocity can be freely chosen as the walls are
frictionless. (ii)~A shear band segment requires a discontinuity of
the tangential velocity with some magnitude $\Delta v$, while the
normal component must be continuous.

An immediate consequence of these two conditions is that shear bands
cannot end at a wall, but must be ``reflected'' under some angle or
end in one of the corners of the rectangular container. Thus, in the
following we consider only shear band configurations represented by
one or more piecewise straight paths ${\cal B}$ that close upon
themselves or begin and end in a corner. Despite these restrictions
the set of admissible shear band configurations is very large. They
will be analyzed in detail in the next sections.

Among these possible configurations we select the ones which minimize
the energy dissipation rate
\begin{equation}
\dot E=\int_{\cal B} \mu \Delta v P dl,
\end{equation}
where the integral extends over the whole path ${\cal B}$ representing
the shear band configuration, $\Delta v$ is the tangential velocity
difference across the shear band, $P$ is the normal stress acting on
the line element $dl$. We suppose that the material is homogeneous and
is described by a single effective friction coefficient $\mu$ which is
constant in the system.

\subsection{Diagonal shear band}

\begin{figure}
\begin{center}
\epsfig{file= 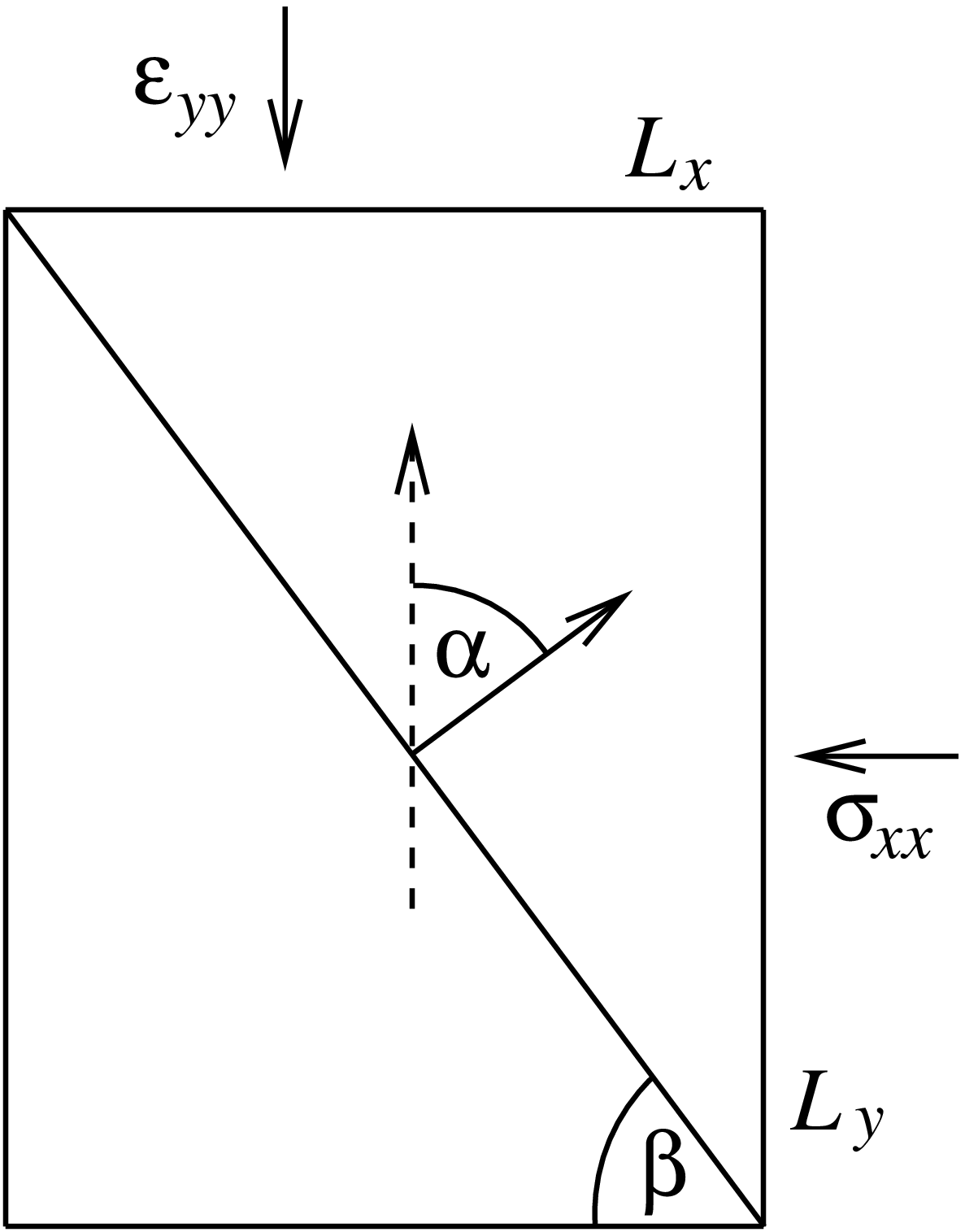,height=3.8cm}
(a)
\quad
\epsfig{file= 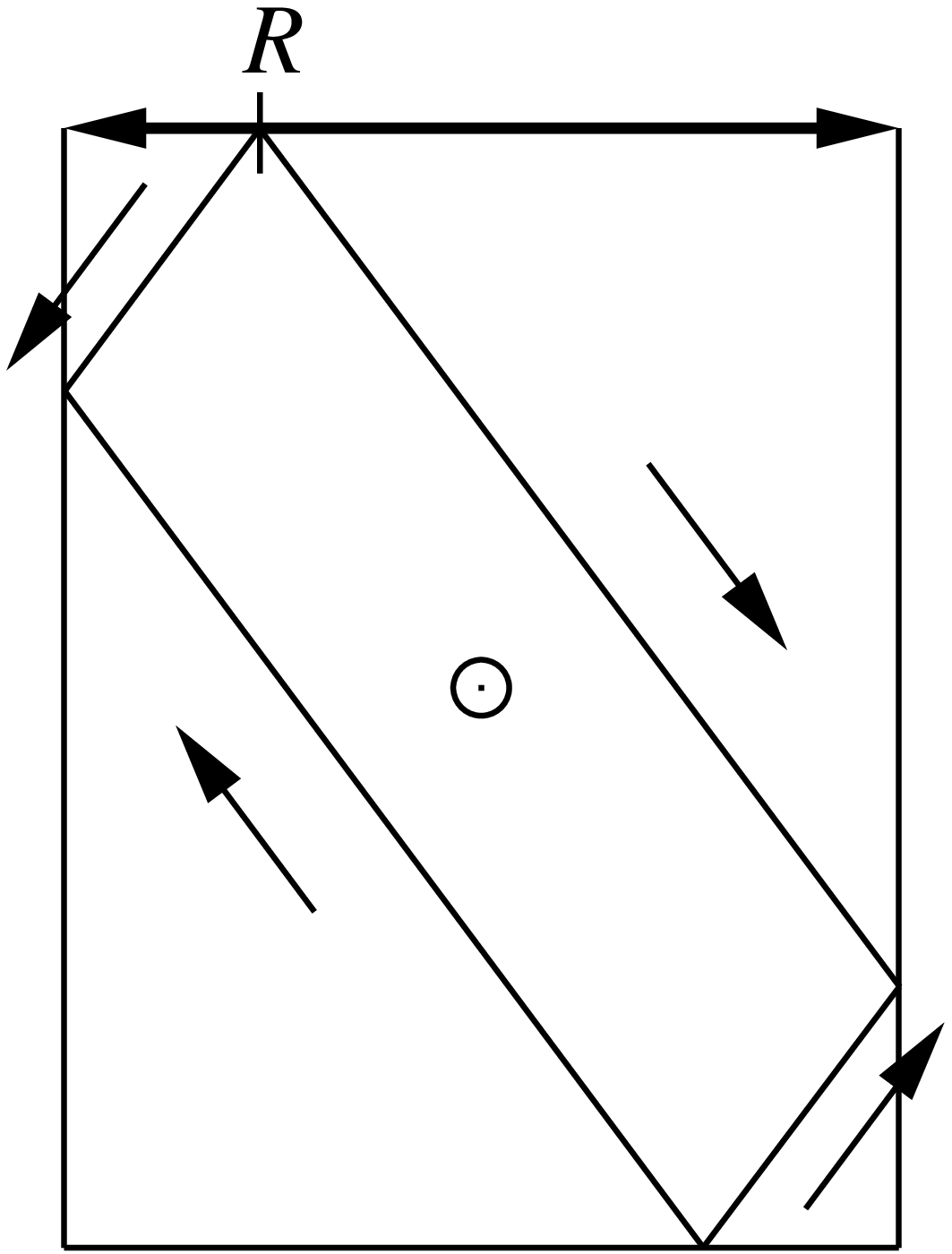,height=3.5cm}
(b)
\caption{\label{Fig_Pressure} (a) The simplest shear band configuration,
the diagonal. (b) Shear band configuration for $\alpha{=}\beta$ with one
degree of freedom.
}
\end{center}
\end{figure}

We start our analysis with the the simplest possible case which is a
single shear band spanning diagonally the sample
[Figs.~\ref{Fig_SBex}(c) and \ref{Fig_Pressure}(a)]. Let us denote the
aspect ratio by $L_y/L_x{\equiv}\tan\beta$, the angle between the major
principal axis (the $y$-axis in the present case) and the normal of
the shear band by $\alpha$ (cf.\ Fig.~\ref{Fig_Pressure}). In this
special case $\alpha{\equiv}\beta$.

The velocity in the two domains should match the bounding walls, so
the velocity difference between the two domains is $\Delta v{=}
u_y/\sin\alpha$. The unit length on the shear band is $dl {=}
dx/\cos\alpha$. $P(\alpha)$ can be derived from force balance on the
shear band as there are no accelerations: the horizontal components of
the normal and tangential force components on the line element are
$P(\alpha) dl \sin\alpha$ and $-\mu P(\alpha) dl \cos\alpha$,
respectively.  Their sum must be equal to $\sigma_{xx} dy$, where
$dy/dl {=}\sin\alpha$, yielding
\begin{equation}
  P(\alpha)=\frac{\sigma_{xx}}{1-\mu\cot\alpha}.
  \label{Eq_NormalStress}
\end{equation}

An immediate consequence is that the \emph{shear band angle} $\alpha$
must be larger than the \emph{internal friction angle}
$\phi\equiv\arctan\mu$, as the normal stress cannot be tensile. When
$\alpha$ approaches $\phi$ the normal stress $P$ diverges, expressing
the impossibility that such a system moves (at finite confining stress
$\sigma_{xx}$). 

Thus the dissipation rate of the diagonal shear band takes the
following form:
\begin{eqnarray}\label{Eq_dissiprate}
\dot E_d(\alpha)&=&\frac{\mu|\eyy|\sigma_{xx}V}
{\cos\alpha\sin\alpha(1-\mu\cot\alpha)}\cr
&=&\mu|\eyy|\sigma_{xx}V\frac{\tan^2\alpha+1}{\tan\alpha-\mu},
\end{eqnarray}
where $V{=}L_xL_y$ denotes the volume of the sample.
Minimizing the above equation yields
\begin{equation}\label{Eq_MC}
\alpha_\mathrm{min}=\amc=\frac{\pi}{4}+\frac{\phi}{2},
\end{equation}
where $\amc$ is the Mohr-Coulomb angle.

The equation (\ref{Eq_MC}) is an important result. It says that the
dissipation rate for the diagonal shear band is minimal when the angle
of the shear band equals to the Mohr-Coulomb angle.

The dissipation rate can be identified with the work per unit
time done by the walls:
\begin{equation}
\dot E=-V\left(\sxx\exx+\syy\eyy\right)
\end{equation}
From this the stress ratio can be derived:
\begin{eqnarray}\label{Eq_stressratio}
s(\alpha)\equiv\frac{\syy}{\sxx}&=&1+\frac{\dot E}{|\eyy|\sxx V}\cr
&=& 1+\frac{\mu}{\cos\alpha\sin\alpha(1-\mu\cot\alpha)}
\end{eqnarray}

The stress ratio of this diagonal configuration varies with the aspect
ratio. Thus the diagonal shear band should be observed only around
$\beta\simeq\amc$ otherwise the stress ratio cannot be constant.

\subsection{Complex shear band structures}

\begin{figure}
\begin{center}
\epsfig{file= 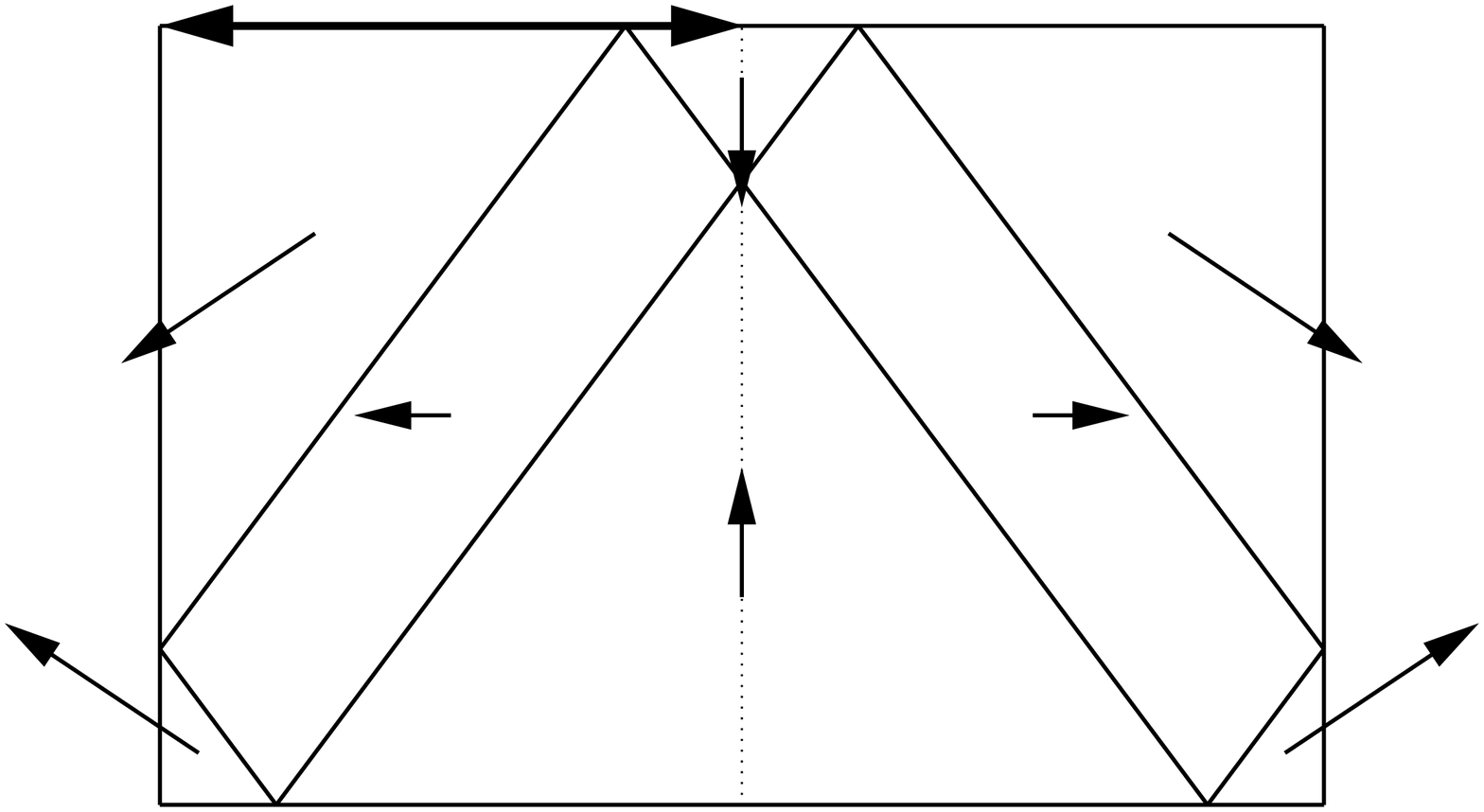,width=4cm}
\epsfig{file= 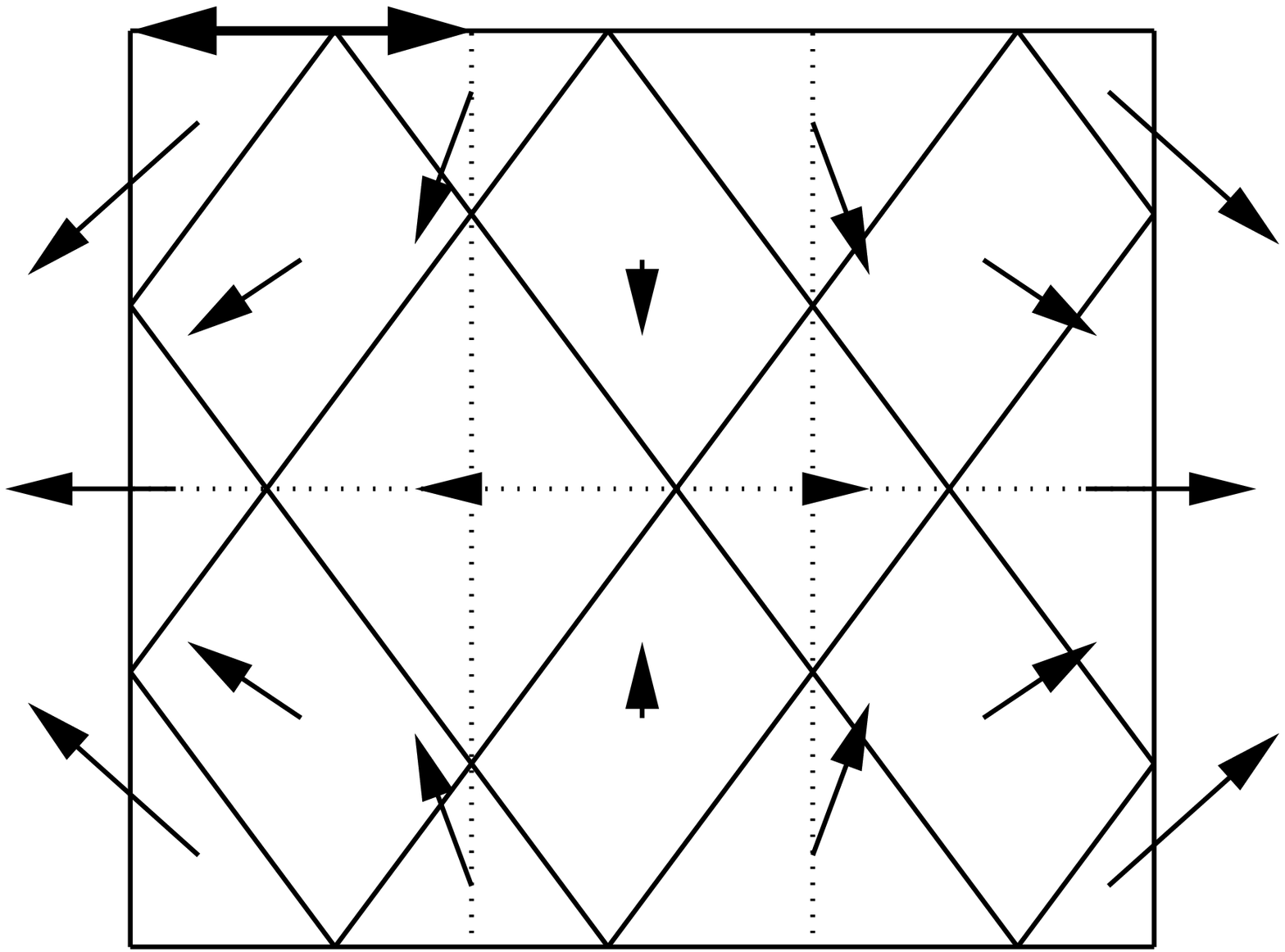,width=4cm}
\linebreak
\null\hfill(a) \hfill\hfill (b) \hfill\null
\caption{\label{Fig_Combine} Combined shear band configuration samples (a) 
$q/r=2/1$ (b) $q/r=3/2$. The arrows show the velocities of the
particles in the domains.
}
\end{center}
\end{figure}

Let us first examine the configuration shown on
Fig.~\ref{Fig_Pressure}(b). For this configuration the formula for the
dissipation rate is the same as in Eq.~(\ref{Eq_dissiprate}) since the
length of the shear band is twice but the velocity jump is half than
for the diagonal shear band. Thus again the minimum of the dissipation
rate is at $\alpha{=}\beta{=}\amc$.

Let us note, that Fig.~\ref{Fig_Pressure}(b) has a degree of freedom.
The reflection point on the top wall can be anywhere. Moving the
reflection point $R$ into the corner reduces this configuration to the
diagonal shear band. Fig.~\ref{Fig_Pressure}(a) is a special case
of Fig.~\ref{Fig_Pressure}(b).

Let us call shear band configuration Fig.~\ref{Fig_Pressure}(b) an
{\em elementary cell}. More complicated shear band configurations can
be constructed if we place $q$ times $r$ elementary cells in a
rectangular lattice by matching the reflection point of the shear
bands and scaling the velocity components to match at the cell
boundaries (examples are shown in Fig.~\ref{Fig_Combine}).  The
minimization presented in \cite{PG09} shows that for these
configurations the dissipation rate is minimal if all shear band
angles are equal and given by $\alpha{=}\amc$. Furthermore the value of
the dissipation rate is the {\em same} as for a single elementary cell
(velocity jump and length scale inversely proportionally), namely:
\begin{equation}
\dot E_\mathrm{min}(\beta)= \dot E_d(\amc)
\end{equation}

It is important to stress that the minimal dissipation rate for the
complex shear band structures is independent of the values of $q$ and
$r$.

On the other hand if $q/r{\not=}1$ then the aspect ratio of the sample
is different from $\tan\amc$. Thus for all aspect ratios $\tan\beta$
and for arbitrary small $\epsilon$ we can create a shear band
configuration with $|q/r-\tan\amc/\tan\beta|{<}\epsilon$ which have all
the same $\dot E_\mathrm{min}$ with shear band angle equaling the
Mohr-Coulomb angle. 

The stress ratio is a linear function of the minimal dissipation rate
cf.~Eq.~(\ref{Eq_stressratio}).
Thus in the scope of this idealized theory at all aspect ratios we
should find a different shear band configuration, and at the same time
observe a constant stress ratio.

This constant stress ratio can be expressed with the effective
friction coefficient $\mu$ cf.~Eqs.~(\ref{Eq_MC}) and
(\ref{Eq_stressratio}):
\begin{eqnarray}
s&=&\left(\mu+\sqrt{1+\mu^2}\right)^2\cr
\mu&=&\tan\arcsin\left(\frac{s-1}{s+1}\right)
\end{eqnarray}

The above formula was already derived in a different way in
\cite{Nedderman}.

The stress ratio can be measured in the simulations to be $s{=}1.54$.
From this value the effective friction coefficient gets $\mu{=}0.213$.

This result answers question (i) of Sec.~\ref{Sec_Shearbandpatterns}.
The apparent macroscopic steady state is the result of the
complete lack of microscopic steady state which at each aspect ratio
favors a different set of shear band configurations with the same
minimal dissipation rate and stress ratio.

The answer for question (ii) is also clear within the scope of these
idealized shear band structures: For all rational
$\tan\amc/\tan\beta$ there are infinitely many pairs of $q$ and $r$
that fulfill $\tan\amc/\tan\beta{=}q/r$, and are equivalent in terms of
dissipation rate. Thus it is impossible to predict which shear band
configuration will be realized.

We should also note here \cite{PG09} that the linear nature of the
system allows for linear combinations of shear band configurations
which also enlarges the set of available configurations for a given
aspect ratio. An example: Fig.~\ref{Fig_SBex} (b) is a linear
combination of Fig.~\ref{Fig_SBex} (c) and its mirror image.

The results shown here are valid for homogeneous materials. In reality
the effective friction coefficient fluctuates around its mean and does
so the dissipation rate. The actual minimum does not only depend on the
average value but also on the variance. The integral in
Eq.~(\ref{Eq_dissiprate}) averages the fluctuations of the effective
friction coefficient along the path. The overall variance of the path
is inversly proportional to the length of the shear band. This way the
shorter shear bands have larger variance and thus more likely to be
minimal.  A consequence is that one can easily find nice, simple shear
band configurations like the examples on Fig.~\ref{Fig_SBex} in almost
all simulations.

In summary in a perfectly homogeneous material one would find
different shear band configurations at each time step. The apparent
macroscopic steady state is the result of the fact that for the
optimal configuration the dissipation rate depends only on the shear
band angle and is independent of the aspect ratio.

\section{Numerical results with frictionless walls}

\subsection{Shear band detection}

An important prediction of our theory is that the shear band angle is
independent of the sample aspect ratio and equals to the Mohr-Coulomb
angle $\amc$. In order to test this result on the simulations we need
to detect the position of the shear band. Many quantities were already
introduced to measure the local shear intensity \cite{Fazekas07}. We
found that for our purpose the best choice is the Frobenius norm of
the gradient of the coarse grained velocity $\vec v$
(see \cite{Goldhirsch01}), which is defined as follows:
\begin{equation}
I_{vg}\equiv||\nabla \vec v||^2=(\partial_x v_x)^2+
(\partial_x v_y)^2+(\partial_y v_x)^2+ (\partial_y v_y)^2
\end{equation}

The strain rate tensor can be decomposed into volumetric
$\dot\varepsilon^V$ and shear $\dot\varepsilon^S$ strain part:
\begin{eqnarray}
\dot\varepsilon_{ij}&=&\frac{\partial_i v_j+\partial_j v_i}{2}=\cr
&=&
\underbrace{\frac{\exx+\eyy}{2}\delta_{ij}}_{\dot\varepsilon_{ij}^V}
+\underbrace{\dot \varepsilon_{ij}-\frac{\exx+\eyy}{2}\delta_{ij}
}_{\dot \varepsilon_{ij}^S},
\end{eqnarray}
which leads to a decomposition of
\begin{equation}
I_{vg}=
2\lambda_\pm^2+
\frac{1}{2}\left(\exx+\eyy\right)^2+
\frac{1}{2}(\nabla\times\vec v)^2,
\end{equation}
where $\lambda_\pm=\pm\sqrt{{\dot\varepsilon^S_{xx}\null}^2+
{\dot\varepsilon^S_{xy}\null}^2}$ are the eigenvalues of the shear
strain tensor.

In this paper we call $I_{vg}$ {\em the local shear intensity}.

\subsection{Shear band direction}

In general the angular correlation of $I_{vg}$ is a good way to
measure the direction of the shear bands marked with high values of
$I_{vg}$. However, in our case both $\alpha$ and $\pi{-}\alpha$ angle
shear bands are possible that cross each other many times. This adds a
lot of noise to the angular correlations. Thus we chose another
method:

In order to measure the directions of the shear bands a segment $\vec
l$ of length $l{=}|\vec l|$ larger than the width of the shear band is
placed at point $\vec x$. Then the correlation between the local shear
intensity is calculated at the end points of the segment {\em only if}
all along the segment $\vec l$ the valus of $I_{vg}$ are larger than
its average $\bar I_{vg}$. Thus the new angular correlation function
is defined as:
\begin{equation}
C_I(\alpha,l)\equiv\langle
I_{vg}(\vec x)I_{vg}(\vec x+\vec l)\rangle_{\vec x,I_{vg}>\bar I_{vg}},
\end{equation}
This definition has the advantage that it avoids contributions from
points lying in different shear band segments.

\begin{figure}
\begin{center}
\epsfig{file= 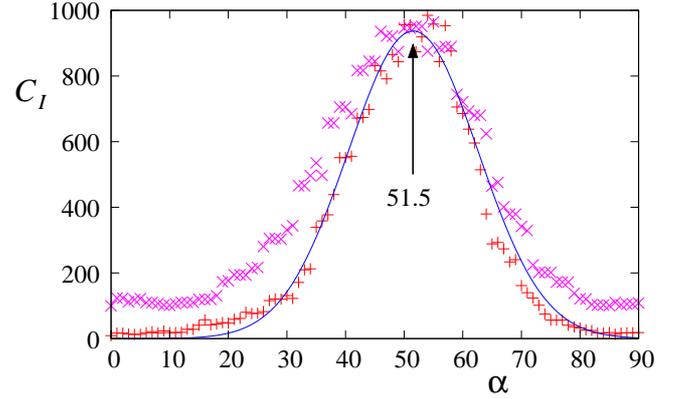,width=8.5cm}
\caption{\label{Fig_ICorr} The segment angular correlation function of
the shear intensity averaged for 4 samples and 400 timesteps, for
$2.6{>}\tan\beta{>}0.5$. The length of the segment was $|\vec l|{=}10$
($\times$), resp. $|\vec l|{=}12$ ($+$) in units of average particle
diameter. The continuous curve is a Gaussian fit to the latter data
for points $C_I{>}500$.
}
\end{center}
\end{figure}

On Fig.~\ref{Fig_ICorr} we show the values of $C_I$ for two
different segment length. It displays a strong peak at an angle of
$\amc{=}51.5^o\pm1^o$. The predicted value from the measured stress
ratio is $\amc{=}51^o$.

As a side result, the distribution functions predict a shear band
width of about $6.5$ particle diameters, which is consistent with the
picture of narrow shear bands.

\section{Frictional walls}

\subsection{The X shaped shear band structure}

The negative answer to the question (ii) of
Sec.~\ref{Sec_Shearbandpatterns} raises another problem of importance.
Namely: Is there a case where the principle of minimal dissipation can
predict the observed shear band configuration? The answer can only be
affirmative, if we can reduce the degeneracy of the possible solutions.

\begin{figure}
\begin{center}
\epsfig{file= 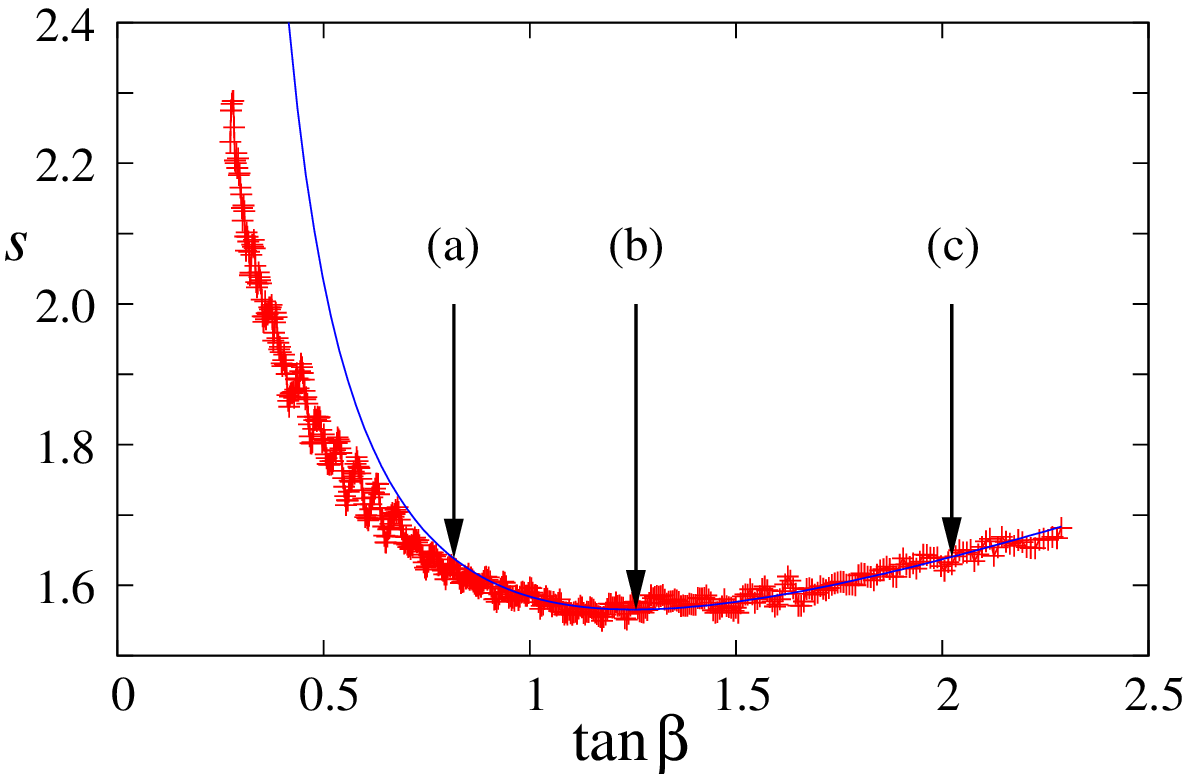,width=8.5cm}
\epsfig{file= 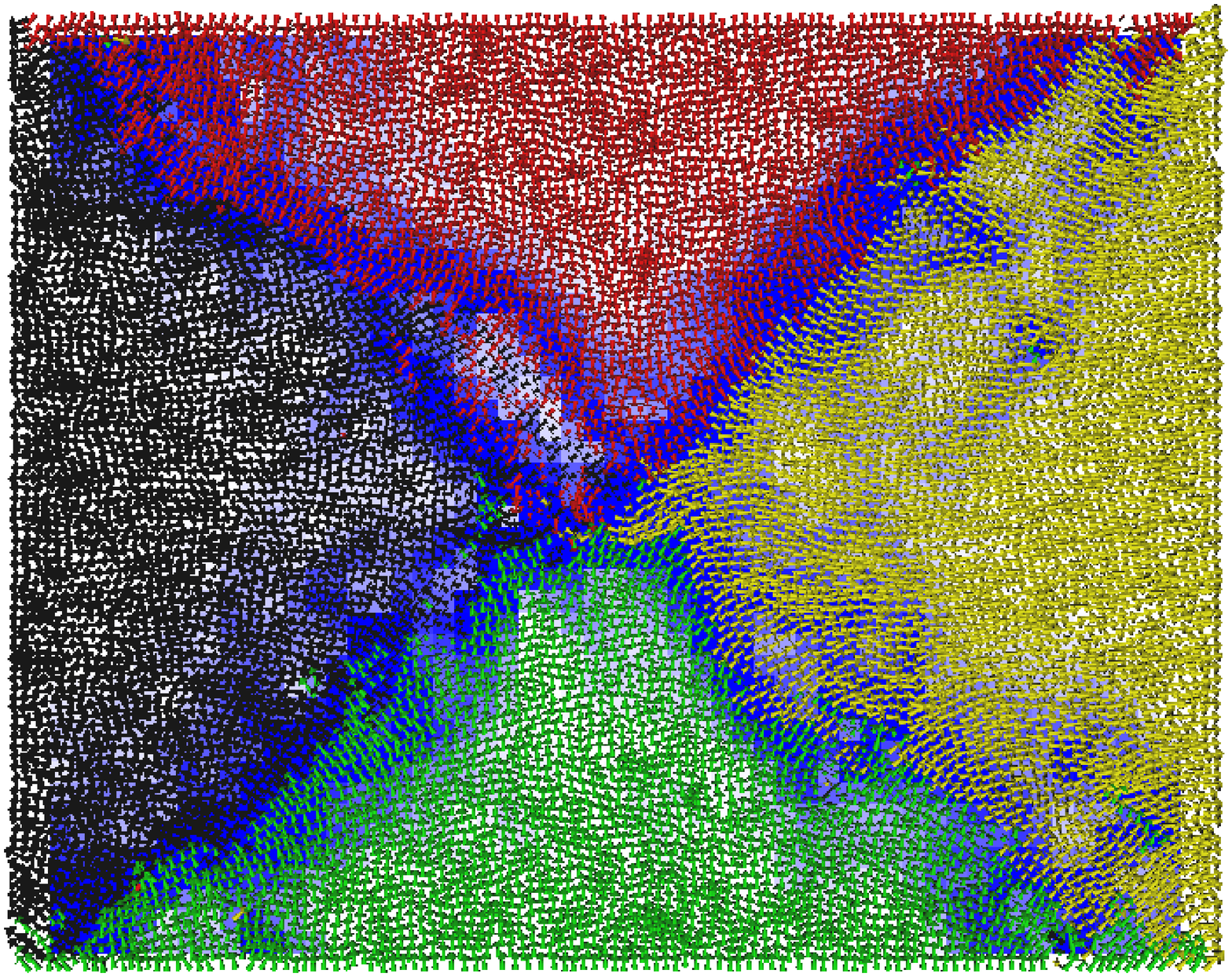,width=3.31cm}
\epsfig{file= 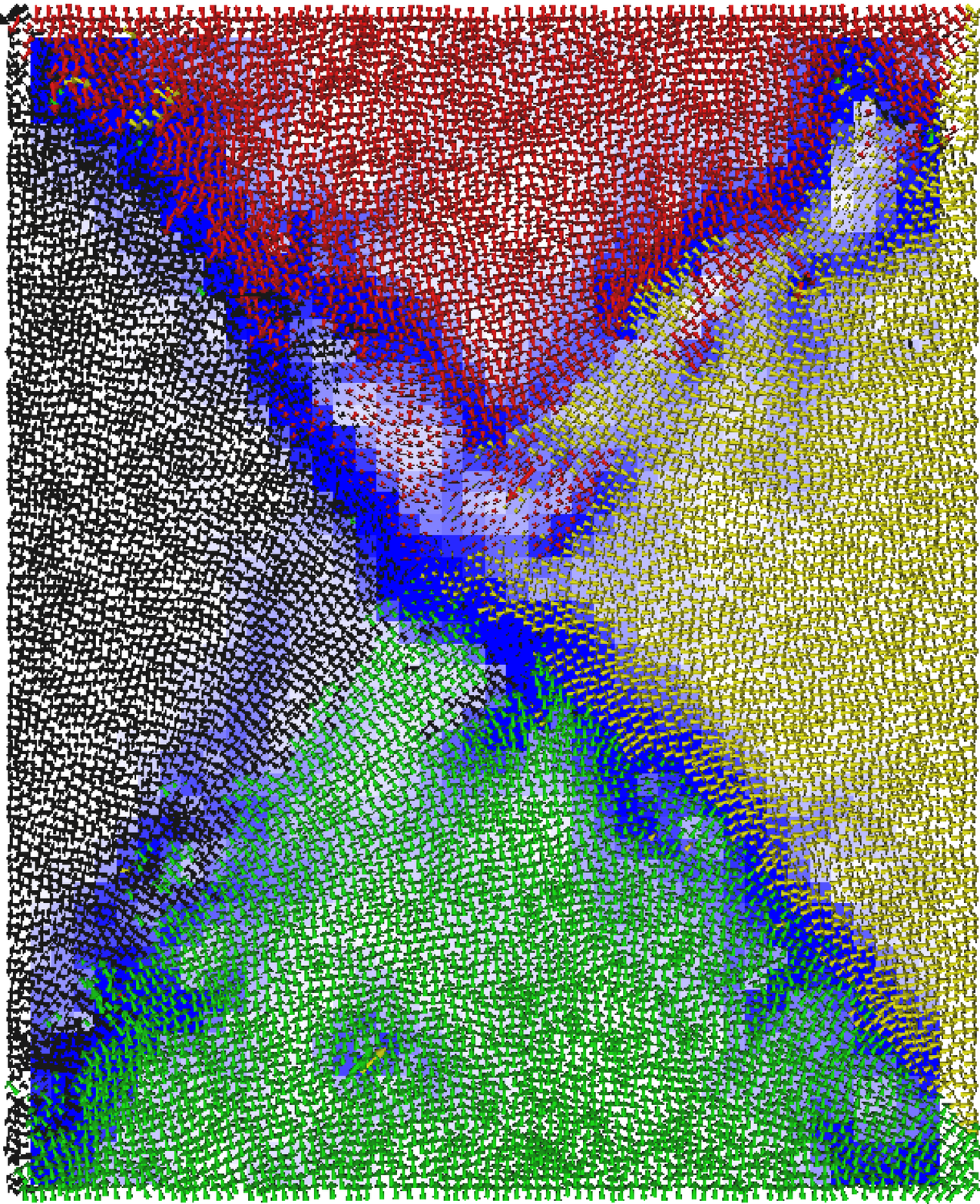,width=2.67cm}
\epsfig{file= 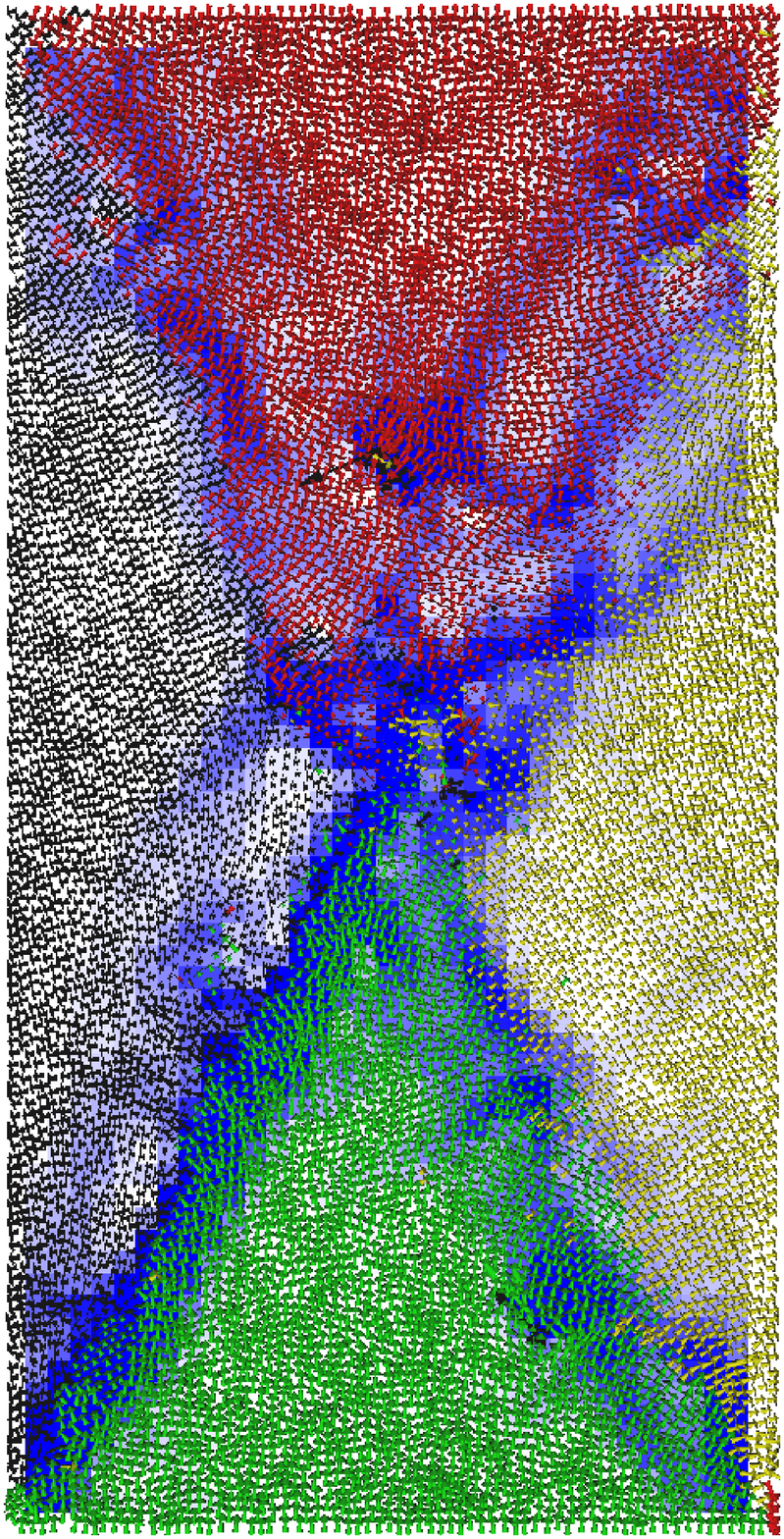,width=1.95cm}
\linebreak
\null\hfill\hfill(a)\hfill\hfill\hfill(b)\hfill\hfill(c)\hfill\null
\linebreak
\caption{\label{Fig_testX} In the top figure the data points show the
stress ratio as function of the angle of the diagonal. The data
points are ensemble averages over 30 different realizations. The
fitted curve is Eq.~(\ref{Eq_stressratio}) with $\mu=0.226$. The snapshots
presented below were taken at the indicated instances. The blue color
is proportional to the shear intensity $I_{vg}$. Particles are 
color coded according to the quarter they are pointing to: red-north,
yellow-east, green-south, black-west.
}
\end{center}
\end{figure}

The simplest way to do this is to introduce friction on the walls.
Wall friction introduces a finite difference in the dissipation rate
for shear band configurations with or without velocity slip at the
walls. Furthermore, it is obvious that there is only one shear band
configuration with no tangential velocity at the walls, since the
following two restrictions must be met: (i) it may not have any shear
band reflected at the walls, which would mean different tangential
velocities in the two domains, (ii) domains touching different walls
should be separated by shear bands. These two conditions imply that
perpendicular velocities in the domains touching the walls are only
possible for the X-shaped shear band configuration, which is composed
of the two diagonal shear bands, see Fig.~\ref{Fig_SBex}(c).

We expect thus that the stress ratio is not constant any more but has
a minimum at aspect ratio of the Mohr-Coulomb angle
($\tan\beta{=}\tan\amc$) which is the optimal for the X-shaped shear
band structure: Furthermore close to this point the stress ratio
should be described by Eq.~(\ref{Eq_stressratio}).

The above equation can be tested on our data. Since $\eyy$ is a
simulation parameter, the only fit parameter we are left with is
the effective friction coefficient of the bulk $\mu$. The value
$\mu=0.226$, Eq.~(\ref{Eq_stressratio}) fits very well the numerical
data over a wide range of aspect ratios as shown on
Fig.~\ref{Fig_testX}. Moreover snapshots of the system also clearly
indicate the X-shaped shear band structure.

The difference between the friction coefficients in  case of
frictionless and frictional walls will be the subject of a future
paper.

The answer to the question raised at the beginning of this section is
thus yes, indeed there are cases when the principle of minimal
dissipation is capable of predicting the actually observed shear
bands. In the next section we discuss if we can predict the overall
stress ratio curve and the other observed shear band configurations.

\subsection{Complex shear band structures}

\begin{figure}
\begin{center}
\epsfig{file= 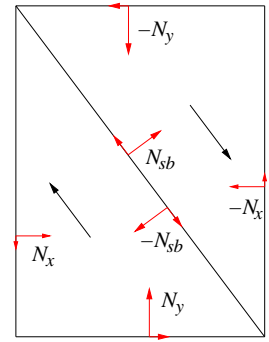,height=4.5cm}
\caption{\label{Fig_frdiagex} The forces acting on the domains in
case of the diagonal shear band.}
\end{center}
\end{figure}

The calculation of the dissipation rate for a general configuration in
case of frictional walls is more complicated because pressure acting
on shear band segments does not only depend on the shear band angle.
Instead it must be calculated from the force balance of the domains.
On Fig.~\ref{Fig_frdiagex} we show an example using the diagonal shear
band.

The force balance shown on Fig.~\ref{Fig_frdiagex} has one external
parameter, the normal force $N_x{=}\sxx L_y$ acting on the confining
walls, two material parameters: the effective friction coefficient of
the bulk $\mu$ and the wall friction $\mu_w$, and two unknowns, the
normal force on the shear band $N_{sb}$ and the normal force on the
compressing walls $N_y{=}\syy L_x$. The force balance of the two domains
gives four (two components for each domain) linear equations for the
unknowns, but for symmetry reasons only two are independent. The
solution of these linear equations allows the calculation of the
dissipation rate and the stress ratio.

The above process sounds simple, but in general it has many problems.
First, in all cases there are more equations than unknowns and it may
happen that a given shear band configuration often seen in the
frictionless simulations gets realizable only at a given aspect ratio.
Second, the analytic solution of the linear system of equations is of
course possible, but the general analytic minimization is impossible
due to the trigonometric functions. So we are left with analyzing
numerically the candidate configurations one by one.

\begin{figure}
\begin{center}
\epsfig{file= 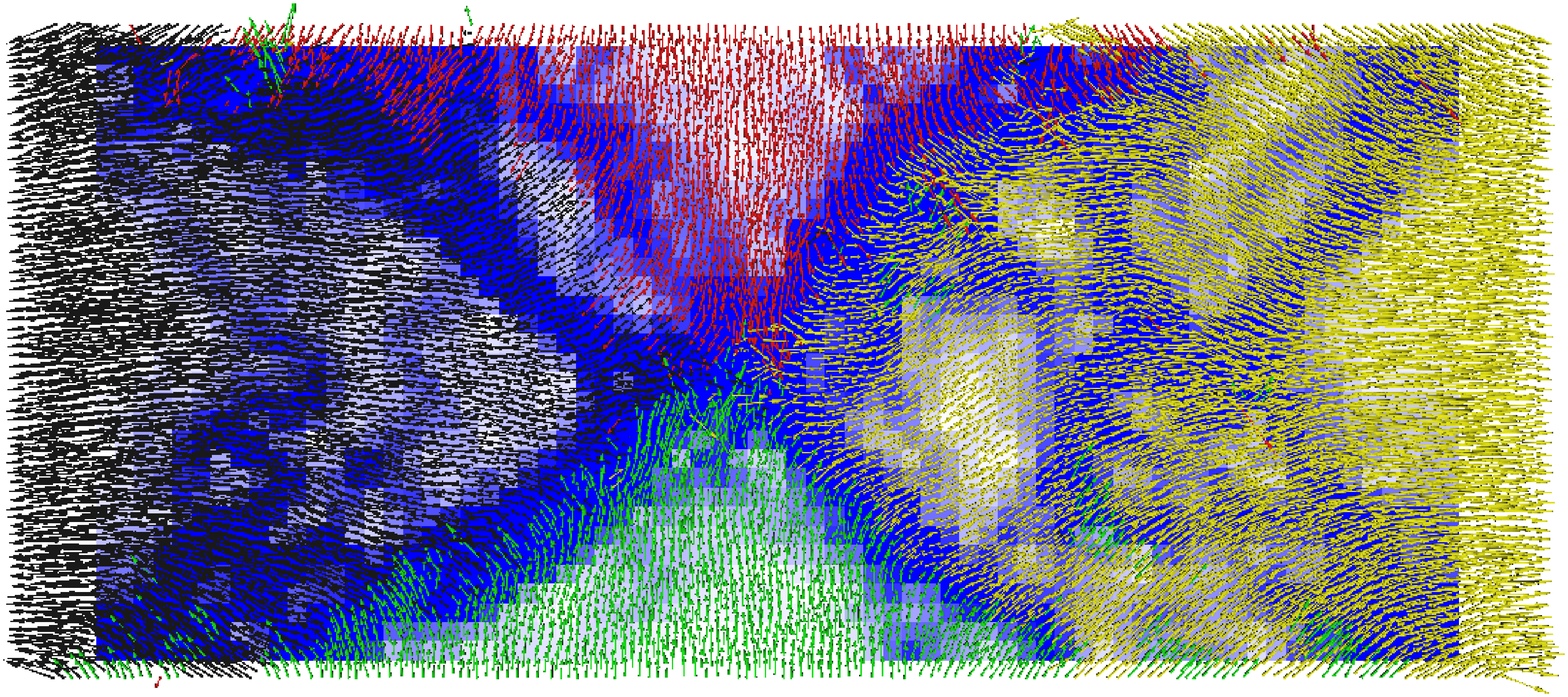,height=3.3cm}
\epsfig{file= 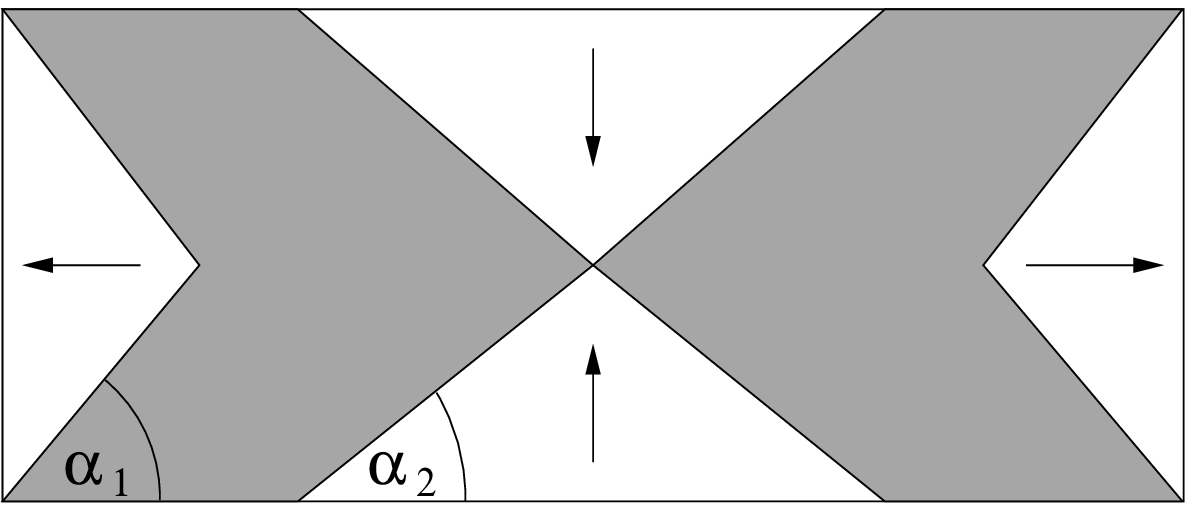,height=3cm}
\caption{\label{Fig_frcomplex} Top: A typical shear band
structure beyond the X shape limit. Bottom: The schematic
representation of the incomplete shear band structure.
}
\end{center}
\end{figure}

Instead of testing this infinite set of configurations - motivated by
the typical shear band structure shown on Fig.~\ref{Fig_frcomplex} -
we define here an {\em incomplete} shear band configuration in which
only part of the shear band system is explicitely defined but still
the minimization can be carried out. Here we focus on the aspect ratio
range of $\beta\ll\amc$ where the X shaped shear band becomes
unfavourable.

Our assumptions are the following: (i) In order to minimize slip on
the walls the generic configuration must have perpendicular velocities
on the largest possible surface: This gives an X shape in the middle
and two half X on the side walls. Since we are in the aspect ratio regime
$\tan\beta{<}1$, the half X shear bands on the confining walls can
cover the whole wall with an optimal angle $\alpha_1{=}\amc$. The
middle X does not cover the whole wall and it has an angle of
$\alpha_2$ which does not necessarily match the Mohr-Coulomb angle.
(ii) The structure in the place left out (gray parts on
Fig.~\ref{Fig_frcomplex}) remains unknown. We know however, that the
force balance is kept on the known shear band segments. (iii) The
velocity slip on the horizontal walls in the gray section is
also unknown. Here we assume a simple linear dependence which gives an
average slip velocity of $u_x/2$. Finally, we note here, that the
visual observations fully support the above assumptions as shown on
Fig.~\ref{Fig_frcomplex}.

\begin{figure}
\begin{center}
\epsfig{file= 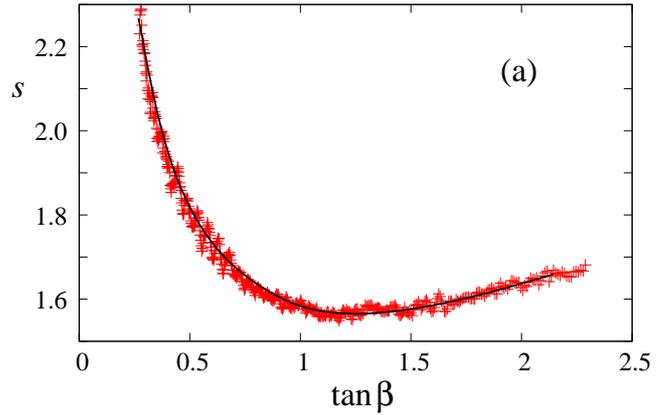,width=8.5cm}
\caption{\label{Fig_frfinalfit} 
Test of the incomplete X configuration on the numerical data.
}
\end{center}
\end{figure}

The minimization with respect to $\alpha_2$ of the dissipation rate
for each aspect ratio is numerically possible if we know the effective
wall friction coefficient. The latter parameter can be determined e.g.
from simple shear simulations \cite{Zahra} and it was found to be
$\mu_w{\simeq}0.12$. The result of numerical minimization at different
aspect ratios match prefectly the simulation data as shown on
Fig.~\ref{Fig_frfinalfit}.

Thus in conclusion the introduction of friction on the walls destroys
the apparent macroscopic steady state but fixes the shear band
structure for a large range of aspect ratios. The prediction of the
stress ratio is still possible with the help of two parameters, the
effective friction of the bulk and the wall. Beside the central regime
we cannot predict the exact shear band structure just some features of
it.

\section{Conclusion}

In conclusion, in this paper we applied the principle of minimal
dissipation to the biaxial shear test. From a theory based on simple
assumptions, in case of frictionless walls, we were able to show that
the apparent macroscopic constant stress ratio is a surprising result
of the permanent minimization of the dissipation rate. This is
manifested by a changing shear band configuration with constant shear
band angle which is equal to the Mohr-Coulomb angle. For each sample
aspect ratios different shear band configurations are optimal, but the
high degeneracy of these optimal configurations denies us the
possibility to predict the observed shear band configurations.

The above degeneracy can be reduced to a single optimal configuration
by introducing friction on the walls, in which case for a wide range
of aspect ratios a single configuration is optimal which can be
clearly observed in simulations. Knowing the optimal configuration, or
at least the important characteristics of it allows us to predict the
the macroscopic stress ratio. 

\section{Acknowledgement}

The partial support by the German-Israeli-Foundation under grant
number I-795-166.10/2003 and by the German Research Foundation via SFB
616 and Piko SPP. Part of this work was funded by the P\'eter
P\'azm\'any program (RET-06/2005) of the Hungarian National Office for
Research and Technology. We also give special thanks to Dirk Kadau who
carried out preliminary simulation work and Tamas Unger for the
discussions.

\end{document}